\documentclass[12pt]{article}
\usepackage{amsfonts}
\usepackage{amsmath}
\usepackage[utf8]{inputenc}
\begin{document}
\newtheorem{theorem}{Theorem}
\newtheorem{lemma}{Lemma}
\newtheorem{example}{Example}
\newtheorem{definition}{Definition}
\newtheorem{corollary}{Corollary}
\newtheorem{proposition}{Proposition}
\newcommand{\ord}{\mathop{\mathrm{ord}}\nolimits}
\newcommand{\Ord}{\mathop{\mathrm{Ord}}\nolimits}
\newcommand{\supp}{\mathop{\mathrm{supp}}\nolimits}
\newcommand{\es}{\mathop{\mathrm{ess\,sup}}\nolimits}
\newcommand{\Id}{\mathop{\mathrm{Id}}\nolimits}
\newcommand{\Tr}{\mathop{\mathrm{Tr}}\nolimits}
\newcommand{\id}{\mathop{\mathrm{Id}}\nolimits}
\newcommand{\Lip}{\mathop{\mathrm{Lip}}\nolimits}
\newcommand{\Qp}{\mathop{\mathbb Q_p}\nolimits}
\newcommand{\Zp}{\mathop{\mathbb Z_p}\nolimits}

\righthyphenmin=2
\title{A note on Decoherence and Fractal Hamiltonians}  
\author{Evgeny Zelenov}
\date{September 20, 2020}
\maketitle

\section{Introduction}

Spin bath model is described by the Hamiltonian
$$
H_{SE} = \frac{1}{2}\sigma_z\bigotimes\underbrace{\sum_{k=1}^{N}e_k\left\{\bigotimes_{i=1}^{k-1} \id^{(i)}\bigotimes\sigma_z\bigotimes_{i=k+1}^{N}\id^{(i)}\right\}}_{H_E}.
$$

Here $\sigma_z$ is the Pauli matrix, $\id$ is the identity operator in 2-dimensional unitary space, $e_k, k=1, 2, \dots, N$ are positive real numbers (coupling constants).

The model was suggested by W.~Zurek \cite{Zurek1} as an environment-induced decoherence model.   It can be interpreted as the interaction of a single spin with a spin bath under the condition of the absence of internal dynamics of the spin and the bath. The central spin couples linearly to each of the environmental bath spin, with strength given by the coupling constants $e_k$. Despite its simplicity, the model has experimental proof. For example, it is capable of explaining the behavior of the Loschmidt echo (see \cite{CDPZ}).

The behavior of the system depends on the coupling constants. W.~Zurek considered the case when coupling constants are bounded, but the sum of its squares tends to infinity as the number of environment spins grows. It was shown that in this case, the fast decoherence of gaussian type occurs.

We are most interesting to analyze the situation when the decoherence does not take place. One of the motivations comes from quantum information theory. Namely, the functioning of a quantum computer is dependent on the ability to maintain quantum coherence.

The paper is organized as follows. In the second section, we briefly describe the evolution of the system and the evolution of the central spin. As the Hamiltonian is diagonal, the model is exactly and simply solvable. In the third section, the decoherence regime is observed based on the papers of W.~Zurek and collaborators. Sections 4 and 5 are devoted to the case of decreasing coupling constants.  In the case of $e_k = 1/\theta^k, \,\theta>2$ the Hamiltonian of the environment  has fractal spectrum. For almost all $\theta$, the decoherence still takes place but seems to be very slow. However, there are some excepting values of $\theta$ when the decoherence does not occur. Namely, it is a fact when $\theta$ is a Pisot number. The proofs are based on known results on Bernoulli convolutions. In section 6, the case of increasing coupling constants is considered. It is shown that if constants satisfy the Weyl growth condition, then the decoherence depends on only the number of spin in the environment and does not depend on time.

\section{Evolution of the system} 

Let the system be in the following pure state at  $t=0$ 
$$
\psi_{SE} = \frac{1}{\sqrt{2}}\left(\phi_++\phi_-\right)\bigotimes\frac{1}{2^{N/2}}\sum_\alpha \phi_\alpha.
$$
Here $\phi_\pm$ are eigenfunctions of $\sigma_z$ and $\phi_\alpha$ are eigenfunctions of the bath Hamiltonian $H_E$.

$$
\phi_\alpha = \bigotimes_{k=1}^N\phi_\pm^{(k)}, \alpha =\underbrace{(+-++\cdots )}_{N} 
$$

Eigenvalues of $H_{E}$ are given by the formula $E_\alpha = \sum_{k=1}^N\pm e_k.$

State of the total system at time $t$ is given by the following expression:
$$
\psi_{SE}(t) = \frac{1}{\sqrt{2}}\phi_+\bigotimes\phi(t) + \frac{1}{\sqrt{2}}\phi_-\bigotimes\phi(-t),
$$
where
$$
\phi(t) = \frac{1}{2^{N/2}}\sum_\alpha\exp\left(-\frac{i}{2}E_\alpha t\right)\phi_\alpha.
$$

Reduced density matrix at time $t$ has the form
$$
\rho_N(t) = \Tr_{H_E}|\phi_{SE}(t)\rangle\langle\phi_{SE}(t)|= \frac{1}{2}
\begin{pmatrix}
1 & r_N(t)\\
r_N(t) & 1
\end{pmatrix}, 
$$
where $r_N(t)$ (decoherence factor) can be expressed in terms of eigenvalues of the bath Hamiltonian
$$
r_N(t) = \frac{1}{2^N}\sum_\alpha \exp\left(-iE_\alpha t\right).
$$
The decoherence factor can be expressed by the following formula:
$$
r_N(t) = \prod_{k=1}^N\cos\left(e_kt\right).
$$

We are interested in the behavior of the decoherence factor for sufficiently large $N$, i.e., in the case of an infinite bath limit (if it exists). In this case, we use the notation

$$
r(t) = \lim_{N\to\infty} r_N(t).
$$

If we have  $r(t)\to 0,\, t\to\infty$ then {\bf decoherence} takes place.

Note that the decoherence factor is the Fourier transform of the infinite Bernoulli convolution (see \cite{sixty}).

\section{Decoherence regime}

\begin{theorem}
Let $D_N^2$ denotes $\sum_{k=1}^Ne_k^2$. If $e_k<C, k=1,2,\dots , N$ and $D_N\to\infty , N\to\infty$ then the following formula is valid:

$$
\lim_{N\to\infty}r_N\left(\frac{t}{D_N}\right) = \exp\left(-\frac{1}{2}t^2\right)
$$
\end{theorem}

Let $\xi_k$ be the random variable with distribution 
$$P_k = \frac{1}{2}\left(\delta (s-e_k)+\delta(s+e_k)\right).$$

Thus $r_N$ is the characteristic function of the distribution the sum of independent random variables $\xi_1+\dots+\xi_N$.

The rest proof follows from the central limit theorem. The Lindeberg condition is valid in this case. The formula above can be interpreted as 

$$
r_N(t)\approx\exp\left(-\frac{1}{2}D_N^2t^2\right)
$$
for sufficiently large $N$. It means that very fast decoherence takes place. This result was obtained by W.~Zurek (\cite{Zurek1}, \cite{Zurek2}, \cite{Schloss}). Contrary to the results mentioned above, we are most interested in the coherence conservation regime.

\section{Coherence conservation regime}

In the case when $\lim_{N\to\infty}D_N = D<\infty$, the Hamiltonian $H_E$ of the bath can be correctly defined in the infinite bath limit $N\to\infty$.

Let $\Omega = \{-1\,,\,1\}^\mathbb N$ be the Cantor cube with the symmmetric Bernoulli measure $\mu$ and $M_\phi, \phi\in L^\infty(\Omega,\mu)$ is the multiplication operator in $L^2(\Omega,\mu)$, $\left(M_\phi f\right)(x) = \phi(x)f(x), \,\,x\in\Omega, \,\,f\in L^2(\Omega, \mu)$. Projector valued measure $P$ is defined by the formula $(P_A\phi)(x) = \chi_A(x)\phi(x), \,A\in\mathcal B(\Omega),\,\phi\in L^2(\Omega,\mu)$. Here $\mathcal B$ is the algebra of Borel sets in $\Omega$, $\chi_A(x)$ is the indicator function of the set $A$. The Hamiltonian $H_E$ of the bath is the multiplication operator $M_F$ where $F$ is defined by the following formula

$$
F(x) = \sum_{k=1}^\infty e_kx_k,\,\Omega\ni x = \{x_1,\,x_2,\,\dots\,x_n\,\dots\},\,x_k=\pm 1.
$$

From the representation above the spectrum of $H_E$ is the set $F(\Omega) = \{F(x), \,x\in\Omega\}$. 

In this case, there exists the limit

$$
r(t) = \lim_{N\to\infty}r_N(t).
$$

Here the function $r(t)$ is continuous positive-definite function and it is (by Bohner's theorem) the Fourier transform of the measure $\nu = \mu\circ F^{-1}$.

Below we are most interested in the case when
$$e_k = \frac{1}{\theta^k}, \,\theta >2.$$

\begin{theorem}
For $\theta >2$ the spectrum of $H_E$ is the $\left(1-\frac{2}{\theta}\right)$-centered Cantor set.
\end{theorem}

Hausdorf dimension of the spectrum of $H_E$ equals $\log 2/\log\theta$.

The statement of the theorem is known in terms of support of the measure $\nu$ (see, for example, \cite{KW}, \cite{Peres}).

\section{Pisot numbers and decoherence}

Pisot number is a real algebraic integer greater than 1, all of whose Galois conjugates are less than 1 in absolute value.

A characteristic property of a Pisot number:  if $\eta  > 1$ is a real number such that the sequence
$\{\{\eta^n\}\}_{n=0}^{\infty}$
measuring the distance from its consecutive powers to the nearest integer is square-summable,  then $\eta$ is a Pisot number. Any integer is a  Pisot number, and any rational not integer is not a Pisot number.

Others examples of Pisot numbers are: platic constant  (positive root of the polinomial $x^3-x-1$), golden section $\frac{1}{2}(1+\sqrt 5)$, silver section $1+\sqrt 2$.

The set of all Pisot numbers is countably infinite.

\begin{theorem}
The decoherence takes place, i.e. $r(t)\to\infty, t\to\infty$, iff $\theta$ is not a Pisot number.
\end{theorem}

This theorem follows from the statement that a $\left(1-\frac{2}{\theta}\right)$-centered Cantor set is a set of uniqueness for the Fourier series iff $\theta$ is a Pisot number (\cite{S}).

For example, for integer $\theta$ decoherence does not take place.

As was mentioned above, rational not an integer is not a Pisot number. It means that in this case, there is decoherence. However, this decoherence is very slow. Namely, the following statement is valid (\cite{K}).

\begin{theorem}
Let $\theta = p/q\in\mathbb Q, \,p\neq 1$ then we have
$$
r(t) = O\left( \left(\frac{1}{\log |t|}\right)^\gamma\right),\, \gamma = -\frac{\log\cos(\pi/2p)}{\log(2\log q/\log p)}>0
$$
\end{theorem}

\section{Big coupling constants}
To complete the picture, let us consider the case of big coupling constants. 

The sequence $\left(e_k\right)_{k=1}^\infty$ satisfies the Weyl growth condition if there exist 
$\epsilon  >0, \,\delta >0$ such that the inequality $|e_n-e_m|>\delta$ holds for all $|n-m|>n/(\log n)^{1+\epsilon}$. For example the sequence $\theta^k,\,|\theta|>1$ is of such type.

The following theorem is valid.

\begin{theorem}
Let the sequence $\left(e_k\right)_{k=1}^\infty$ of coupling constants satisfy the Weyl growth condition.

Then the following formula is valid for almost all $t\in\mathbb R$:
$$
\lim_{n\to\infty} \frac{1}{n}\log |r_n(t)| = -\log 2.
$$
\end{theorem}

It means that for sufficiently large $N$ we have $|r_N(t)|\approx 2^{-N}$ independently on $t$. That is, the system is ``frozen'' in a quantum state, and the decoherence does not take place.

Proof is based on the following statement (see, for example, \cite{KuipNied}): 
if the sequence $(x_n)_{n=1}^\infty$ is uniformly distributed $\mod 1$, then for all continuous function $f,\, f\in C[0,1]$ we have
$$
\lim_{n\to\infty}\frac{1}{n}\sum_{i=1}^{n}f\left(\{x_n\}\right) = \int_0^1f(x)dx.
$$
In the last formula, $\{x\}$ denotes the fractional part of $x$.

 If the sequence of coupling constants satisfies the Weyl growth condition then the sequence $\left(e_n t/\pi\right)_{n=1}^\infty$ is uniformly distributed $\mod 1$ for almost all $t\in\mathbb R$ (see \cite{KuipNied}). Therefore we have

 $$
 \lim_{n\to\infty}\frac{1}{n}\log |r_n(t)| =  \lim_{n\to\infty}\frac{1}{n}\sum_{i=1}^n\log|\cos e_it| = \int_0^1\log|\cos\pi x|dx = -\log 2.
 $$

 The idea to use uniformly distributed sequences to prove the theorem borrowed from \cite{sixty}.

\end{document}